\documentclass[twocolumn]{aastex63}

\usepackage[version=4]{mhchem}
\usepackage{multirow}
\received{}
\revised{}
\accepted{}
\submitjournal{ApJL}

\shorttitle{No clear relation between diffusion energy and desorption energy}
\shortauthors{Furuya et al.}

\newcommand{\esd}{E_{\rm sd}}
\newcommand{\edes}{E_{\rm des}}

\begin{document}

\title{Diffusion activation energy and desorption activation energy for astrochemically relevant species on water ice show no clear relation}

\correspondingauthor{Kenji Furuya}
\email{kenji.furuya@nao.ac.jp}

\author[0000-0002-2026-8157]{Kenji Furuya}
\affiliation{National Astronomical Observatory of Japan, Osawa 2-21-1, Mitaka, Tokyo 181-8588, Japan}
\author[0000-0002-4991-4044]{Tetsuya Hama}
\affiliation{Komaba Institute for Science, The University of Tokyo, Meguro, Tokyo 153-8902, Japan}
\author[0000-0002-6852-3604]{Yasuhiro Oba}
\affiliation{Institute of Low Temperature Science, Hokkaido University, Sapporo, Hokkaido 060-0819, Japan}
\author[0000-0002-0495-5408]{Akira Kouchi}
\affiliation{Institute of Low Temperature Science, Hokkaido University, Sapporo, Hokkaido 060-0819, Japan}
\author[0000-0001-8408-2872]{Naoki Watanabe}
\affiliation{Institute of Low Temperature Science, Hokkaido University, Sapporo, Hokkaido 060-0819, Japan}
\author[0000-0003-3283-6884]{Yuri Aikawa}
\affiliation{Department of Astronomy, The University of Tokyo, Tokyo, 113-0033, Japan}



\begin{abstract}
The activation energy for desorption ($\edes$) and that for surface diffusion ($\esd$) of adsorbed molecules on dust grains are two of the most important parameters for the chemistry in the interstellar medium.
Although $\edes$ is often measured by laboratory experiments, the measurement of $\esd$ is sparse.
Due to the lack of data, astrochemical models usually assume a simple scaling relation, $\esd = f \edes$, 
where $f$ is a constant, irrespective of adsorbed species.
Here, we experimentally measure $\esd$ for \ce{CH4}, \ce{H2S}, OCS, \ce{CH3OH}, and \ce{CH3CN} on water-ice surfaces using an ultra-high-vacuum transmission electron microscope (UHV-TEM).
Compiling the measured $\esd$ values and $\edes$ values from the literature, 
we find that the value of $f$ ranges from $\sim$0.2 to $\sim$0.7, depending on the species.
Unless $f$ (or $\esd$) for the majority of species is available, a natural alternative approach for astrochemical models is running multiple simulations, varying $f$ for each species randomly.
In this approach, ranges of molecular abundances predicted by multiple simulations, rather than abundances predicted by each simulation, are important.
We here run 10,000 simulations of astrochemical models of molecular clouds and protostellar envelopes, randomly assigning a value of $f$ for each species.
In the former case, we identify several key species whose $\esd$ most strongly affects the uncertainties of the model predictions; 
$\esd$ for those species should be investigated in future laboratory and quantum chemical studies.
In the latter case, uncertainties in the $\esd$ of many species contribute to the uncertainties in the model predictions.
\end{abstract}





\section{Introduction}
\label{sec:intro}
Various molecules have been detected in the interstellar medium (ISM) 
both in the gas phase and in the solid phase (i.e., ice). 
The most abundant molecules, such as \ce{H2O} and \ce{CO2}, are formed as ice on grain surfaces via the Langmuir–-Hinshelwood mechanism, where precursor species are adsorbed onto the surface, 
diffuse across the surface, and react when they encounter each other \citep[e.g.,][]{hama13}.
The grain-surface chemistry affects the gas-phase composition through thermal and non-thermal desorption \citep[e.g.,][]{jorgensen20}.
Extensive effort has been devoted to understanding the gas and ice chemistry observed in the ISM through laboratory experiments, quantum chemistry calculations, and astrochemical simulations, where the first two can provide input parameters for the last \citep[e.g.,][]{cuppen17}.


The most fundamental parameters to describe the surface chemistry in the ISM are the activation energy for desorption ($\edes$) from dust grains and that for surface diffusion ($\esd$) on dust grains.
The thermal desorption ($k_{\rm des}$) and thermal diffusion rates ($k_{\rm sd}$) are given by
\begin{align}
k_{\rm des} = \nu_{\rm des} \exp(-\edes/kT), \\
k_{\rm sd} = \nu_{\rm sd} \exp(-\esd/kT), \label{eq:ksd}
\end{align}
respectively, where $\nu$ is a pre-exponential factor and $k$ is the Boltzmann constant.
Numerous laboratory measurements of $\edes$ for stable molecules on astrophysically relevant surfaces has been conducted using the temperature-programmed desorption (TPD) technique \citep[e.g.,][]{collings04,noble12,he16a,smith16,minissale22}.
However, measuring $\esd$ without employing models is more difficult, and has only been successful for atomic species \citep{watanabe10,hama12,minissale15,minissale16}.
Some laboratory studies to measure $\esd$ using infrared spectroscopy of mixed or layered ice of water and targeted molecules have been reported \citep[e.g.,][]{oberg09,Karssemeijer14,he18,mate20}. 
These previous studies constrained $\esd$ on porous amorphous solid water (p-ASW), where Fick's law of diffusion was adopted and the targeted molecules were assumed to diffuse at the surface of pores and/or cracks in p-ASW, but not in the bulk ice.
\citet{mate20} have claimed, however, that diffusion parameters obtained on the basis of Fick's law (i.e., macroscopic diffusion) differ from those for each thermal hop, the latter of which are more relevant in astrochemical models.
Indeed, microscopic Monte-Carlo simulations of \ce{CH4} diffusion on p-ASW with the \ce{CH4} diffusion rates constrained by the experiments with Fick's law could not reproduce experimental results \citep{mate20}.
Because $\esd$ is poorly quantified,
astrochemical models usually assume that the $\esd$ for each chemical species takes a universal, fixed ratio with respect to $\edes$ \citep[e.g.,][]{cuppen17}.

Recently, \citet{kouchi20} developed a new method to measure $\esd$ directly.
They observed the deposition of CO (and \ce{CO2}) on p-ASW in situ using an ultrahigh-vacuum transmission electron microscope (UHV-TEM).
When the substrate temperature was high enough for surface diffusion of CO, the formation of crystalline islands of CO was observed.
During CO deposition, the number of crystalline islands increased, eventually reaching saturation.
At that point, the timescale for CO to diffuse the mean distance between the islands should be equal to the timescale for CO adsorption.
By repeating the experiments at different substrate temperatures, 
they derived the $\esd$ for CO on p-ASW.
In this work, we perform experiments similar to those of \citet{kouchi20} but for \ce{CH4}, \ce{H2S}, OCS, \ce{CH3OH}, and \ce{CH3CN} on compact ASW (c-ASW), and constrain the $\esd$ for these species.

This is the first work that constrains $\esd$ values for multiple species by a single experimental method without employing models, 
and we clearly shows that the ratio of $\esd/\edes$ ($f$) is not universal.
Upon the experimental results, we perform simulations to evaluate the impact of changing f values 
on the abundances of important molecules in models. 



\section{Experiments} \label{sec:experiment}
\subsection{Methods}
Experiments were performed using a UHV-TEM.
Details of the UHV-TEM are described elsewhere \citep{kouchi21FrCh}.
Briefly, the column of the electron microscope was evacuated to $\sim$1$\times10^{-6}$ Pa using five ion pumps.
The pressure near the specimen was kept lower than $\sim$1$\times10^{-6}$ Pa, because the specimen was surrounded by a liquid-\ce{N2} shroud.
We used a Gatan ULTST cooling holder with liquid He for the sample cooling of \ce{CH4}, \ce{H2S}, and OCS, and liquid \ce{N2} for those of \ce{CH3CN} and \ce{CH3OH}.
Five-nonometer-thick amorphous Si and SiN films (SiMPore, US100-A05Q33 and SN100-A05Q33A, respectively) were used as the substrates for sample deposition.
One of the three ICF 70 ports, which were directed at the specimen’s surface at an incidence angle of 55$^{\circ}$, was used for sample deposition. 
A Ti gas inlet tube with a 0.4 mm inner diameter was faced toward the substrate.
Deposition experiments were performed following the protocol developed by \citet{kouchi20,kouchi21}.
Briefly, thin films of $\sim$10-nm thickness c-ASW were prepared by depositing ASW with a deposition rate of $\sim$0.06 nm/s at $\sim$80 K.
Then the films were heated to specific temperatures with a heating rate of $\sim$0.07 K/s, followed by 5-min annealing at $\sim$100 K for \ce{CH4}, \ce{H2S}, and OCS experiments and at 130 K for \ce{CH3CN} and \ce{CH3OH} experiments.
The higher annealing temperature of 130 K was adopted for \ce{CH3CN} and \ce{CH3OH}, 
because the experiments for these molecules were performed at temperatures greater than 100 K.
\citet{kimmel01} found that the surface areas of $\sim$20-nm ASW annealed at 100 K and 130 K are different ($\sim$12 ML versus $\sim$5 ML) in their experimental conditions.
However, we consider that the difference in annealing temperature could have no significant effect on the diffusion in our experiments; surface areas of ASW annealed at 100 K and 130 K in our experiments could become similar, because the thickness of ASW is thinner, the heating rate is much slower, and the annealing time is much longer compared with those in \citet{kimmel01}.

We then deposited \ce{CH4} (99.999 \%, Takachiho Chemical Industrial), \ce{H2S} ($>$99.9 \%, Sumitomo Seika Chemicals), OCS (99.9 \%, Taiyo Nippon Sanso), \ce{CH3CN} ($>$99.9 \%, Sigma-Aldrich), or \ce{CH3OH} (99.5 \%, Kishida Chemical) onto c-ASW for 4 min at a deposition rate of $\sim$1 nm min$^{-1}$. 
\citet{kouchi20,kouchi21} observed CO and \ce{CO2} deposition in situ throughout the deposition process using UHV-TEM.
However, in the present study, because we found that the long electron irradiation of all the adsorbed molecules creates products, such as \ce{H2SO4} and amorphous sulfur in the case of OCS, despite the use of a low-intensity electron beam ($\sim 6 \times10^{-3}$ electrons \AA$^{-2}$ at the sample position), we only observed after deposition by applying low--dose technique \citep{tachibana17}.
Specifically, the accelerating voltage was 80 kV, a very low-intensity electron beam ($\sim 6 \times10^{-3}$ electrons \AA$^{-2}$) was used, and the observations were conducted at low magnification (20,000$\times$).


\begin{figure*}[ht]
\plotone{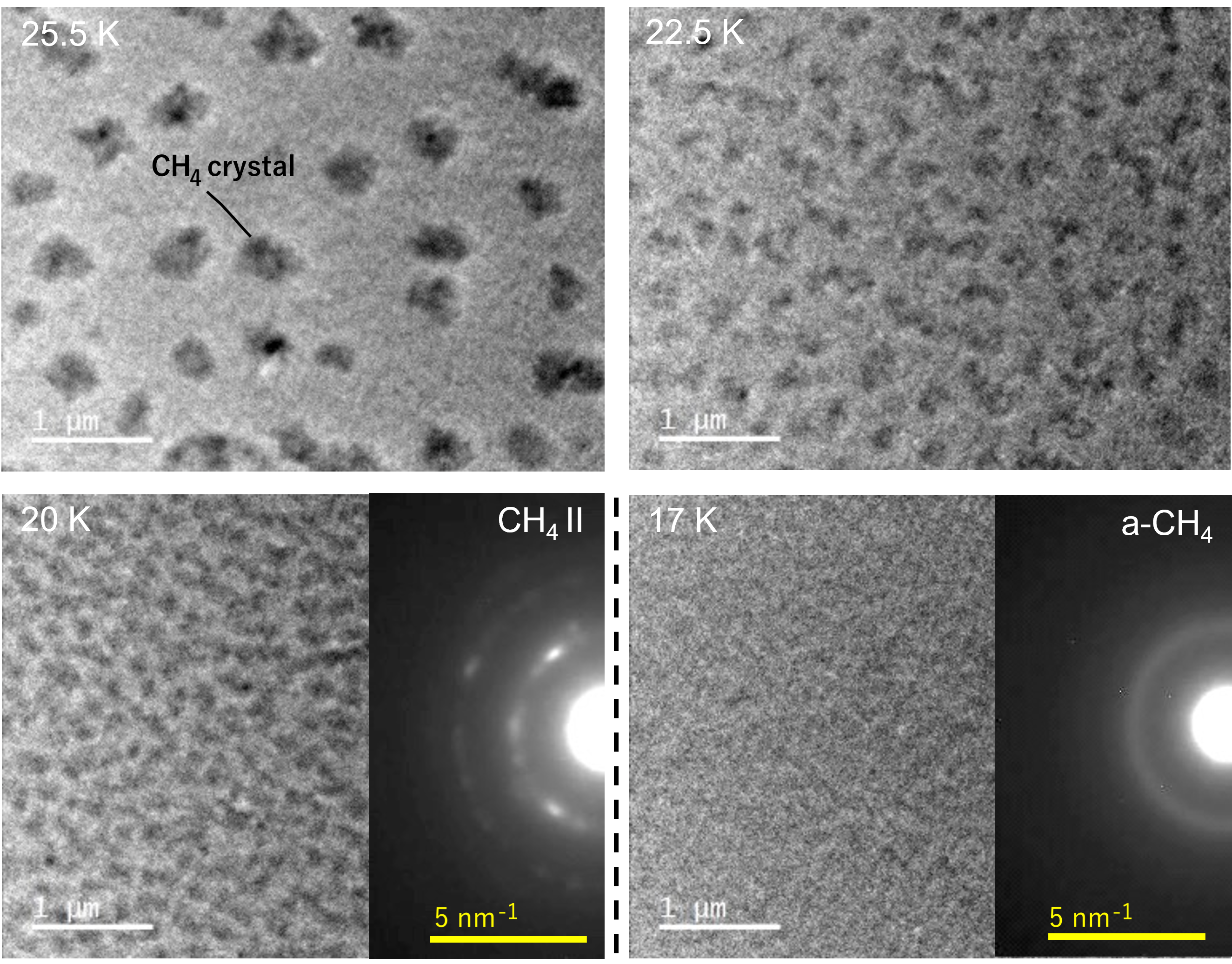}
\caption{
TEM images of \ce{CH4} deposited on on c-ASW at different substrate temperatures. The corresponding electron diffraction patterns are shown in some panels.
\ce{a-CH4} indicates amorphous \ce{CH4}.
The contrast in the TEM images increases from bright (gray) to dark (black) with increasing atomic number (scattering contrast). The contrast for crystalline samples is much stronger (darker) than that for the amorphous sample because of diffraction contrast. Broken lines represent critical temperatures for the formation of crystalline or amorphous solids.
}
\label{fig:tem}
\end{figure*}

\subsection{Results}
\begin{figure*}[ht]
    \plotone{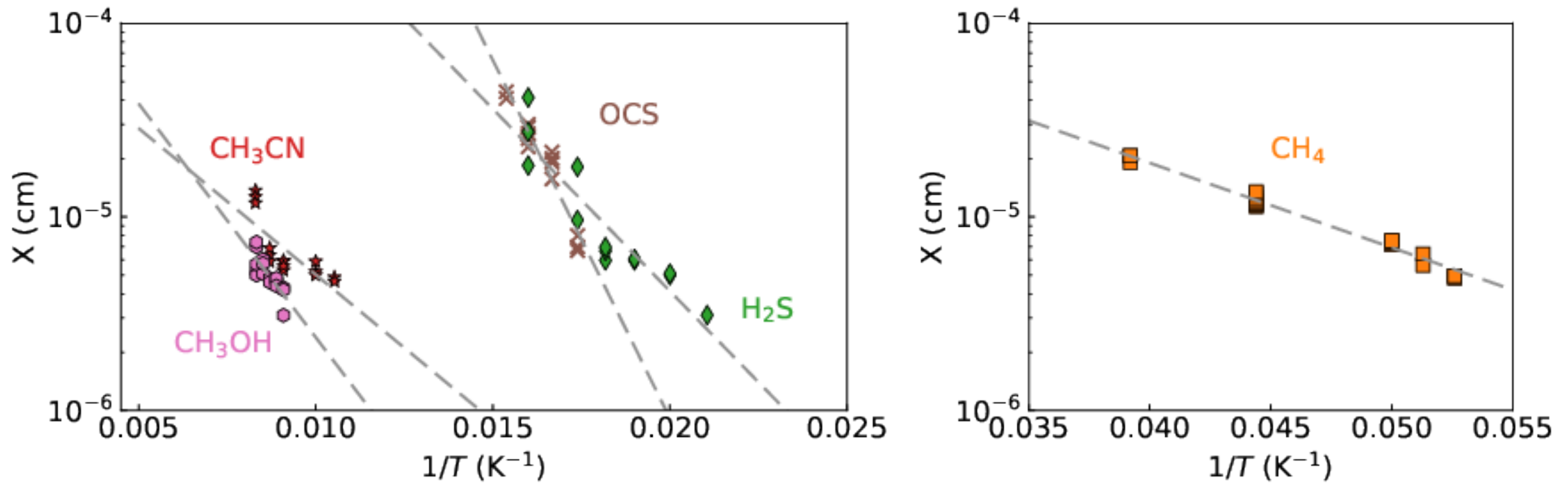}
    \caption{Mean diffusion distance $X$ for the deposited molecules on c-ASW as a function of the inverse of the substrate temperature.
    Gray dashed lines represent least-squares fits.}
    \label{fig:diff_dist}
\end{figure*}

Figure \ref{fig:tem} shows the temperature dependence of the TEM images of deposited samples of \ce{CH4} on c-ASW (see Figures \ref{fig:tem2}--\ref{fig:tem5} in the Appendix A for TEM images of the other molecules).
With decreasing substrate temperature, the number density of crystalline islands increased for all the investigated molecules, and the crystalline sizes decreased.
The crystal structures were determined from the electron diffraction patterns for \ce{CH4} II \citep{press72}, \ce{H2S} III \citep{cockcroft90}, OCS I \citep{overell82}, $\beta$-\ce{CH3CN} \citep{enjalbert02}, and $\alpha$-\ce{CH3OH} \citep{torrie89}.
At lower temperatures indicated by broken lines in the respective figures, formation of uniform amorphous films, as indicated by the halo patterns in the electron diffraction patters, wes observed.
The transition temperatures between the crystalline and amorphous solids of \ce{CH4}, \ce{H2S}, OCS, \ce{CH3CN} and \ce{CH3OH} 
were $\sim$18 K, $\sim$46 K, $\sim$56 K, $\sim$90 K, and $\sim$100 K, respectively.
From the number density of crystalline islands, the mean diffusion distance $X$ for the molecules on c-ASW was derived by the method of \citet[][see also the Appendix A]{kouchi20}.
Figure \ref{fig:diff_dist} shows a plot of $\ln X$ versus the inverse of the temperature;
the slope corresponds to $-\esd/2k$ \citep{kouchi20}.

\begin{figure*}[ht]
    \plotone{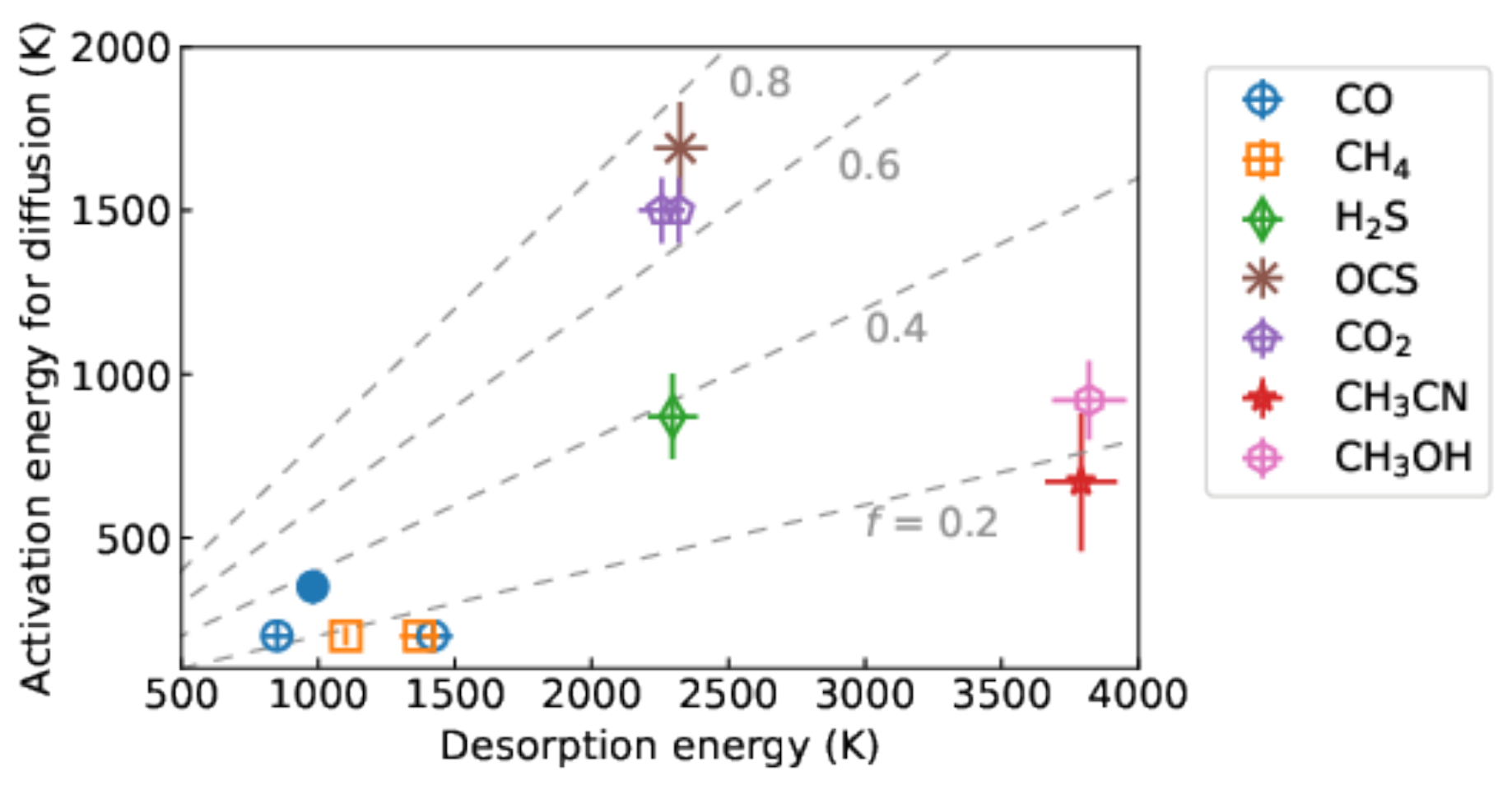}
    \caption{
    $\edes$ versus $\esd$ for 
    various molecules on c-ASW. The crosses inside each symbol indicate error bars. Gray lines indicate $f=0.2$, 0.4, 0.6, or 0.8.
    Filled blue circle indicates CO on p-ASW for comparisons.}
    \label{fig:esd_edes}
\end{figure*}

Figure \ref{fig:esd_edes} shows the $\esd$ measured by UHV-TEM in this work and in our previous work \citep{kouchi20,kouchi21} versus $\edes$ reported in the literature 
(see Table \ref{table:esd_edes} in the Appendix A for values and references).
No clear relation is observed between $\esd$ and $\edes$, and the diffusion-to-desorption activation energy ratio $f$ ranges from 0.14 to 0.73.
This result clearly contradicts the widely applied assumption in astrochemical models that $f$ has a universal, constant value for all species.


\section{Uncertainties in astrochemical models} \label{sec:discuss}
Because there is no universal value for $f$, 
it is inappropriate to use the constant $f$ value in astrochemical models.
Unless $f$ (or $\edes$) for all species is quantified via laboratory experiments or quantum chemical calculations, a natural alternative way is to run multiple simulations, varying the value of $f$ for each species randomly.
In this approach (hereafter statistical approach), ranges of molecular abundances predicted by multiple simulations, rather than abundances predicted by each simulation, are important.


%

There are only a few previous works that employed the statistical approach to investigate the uncertainties in model predictions.
\citet{penteado17} investigated the effect of uncertainties in $\edes$ 
on the prediction of pseudo-time-dependent (i.e., physical conditions are fixed with time) astrochemical models of dense molecular clouds; 
gas and dust temperatures are 10 K, gas density is $2\times10^4$ cm$^{-3}$, 
and the visual extinction ($A_V$) is 10 mag.
They randomly chose $\edes$, considering recommended values and their uncertainties based on a literature search.
The $f$ value was fixed at 0.3 for all species.
They found that $\edes$ for atomic C, HCO, HNO, and \ce{CH2} have the greatest influence on the final ice abundances.
Because $\esd$ was inferred from $\edes$ and because thermal desorption is negligible for these species at 10 K, these dependencies most likely involve diffusion rates \citep{penteado17}.
\citet{iqbal18} investigated the effect of uncertainties in $\esd$ on 
the prediction of astrochemical models for dense molecular clouds.
In their study, $f$ for each species was varied randomly in the range between 0.25 and 0.75, except that for atomic H, whose $\esd$ was fixed.
$\edes$ was fixed at a recommended value for each species.
They found that when the uncertainties in $f$ for several key species,
including \ce{H2}, CO, atomic O, and atomic N, were limited,
the uncertainties in the abundances of almost all species in their model were eliminated.
The difference in the key species identified by \citet{penteado17} and \citet{iqbal18} seems to come from, at least in part, the difference in adopted $\edes$.
For example, \citet{iqbal18} set $\edes$ for atomic C to 10,000 K \citep{wakelam17}, 
whereas \citet{penteado17} varied it in the range of $715 \pm 360$ K.
These previous studies have shown that $\esd$ for several key species determines the uncertainties of model predictions under static physical conditions appropriate for dense molecular clouds.

Here, we expand the studies by \citet{penteado17} and \citet{iqbal18} to more realistic time-dependent physical conditions:
the formation and early evolutionary stage of molecular clouds and infalling envelopes around a low-mass protostar.
Our molecular cloud model (hereafter "model MC") and protostellar envelopes model (hereafter "model PE") are similar to those presented in \citet{furuya15} and in \citet{furuya16}, respectively;
however, here we run each model 10,000 times, 
randomly varying $f$ for each surface species, with $\edes$ fixed.
We address the following three questions:
(i) How do the uncertainties in $\esd$ affect the abundances predicted by astrochemical models?,
(ii) Are there key species that govern uncertainties in the predicted abundances?,
(iii) How well can the conventional method, where $f$ is assumed to be universal for all species and is treated as a free parameter, evaluate the abundance uncertainties due to the uncertainties in $\esd$?

\subsection{Model description}
\begin{figure*}[t]
    \plotone{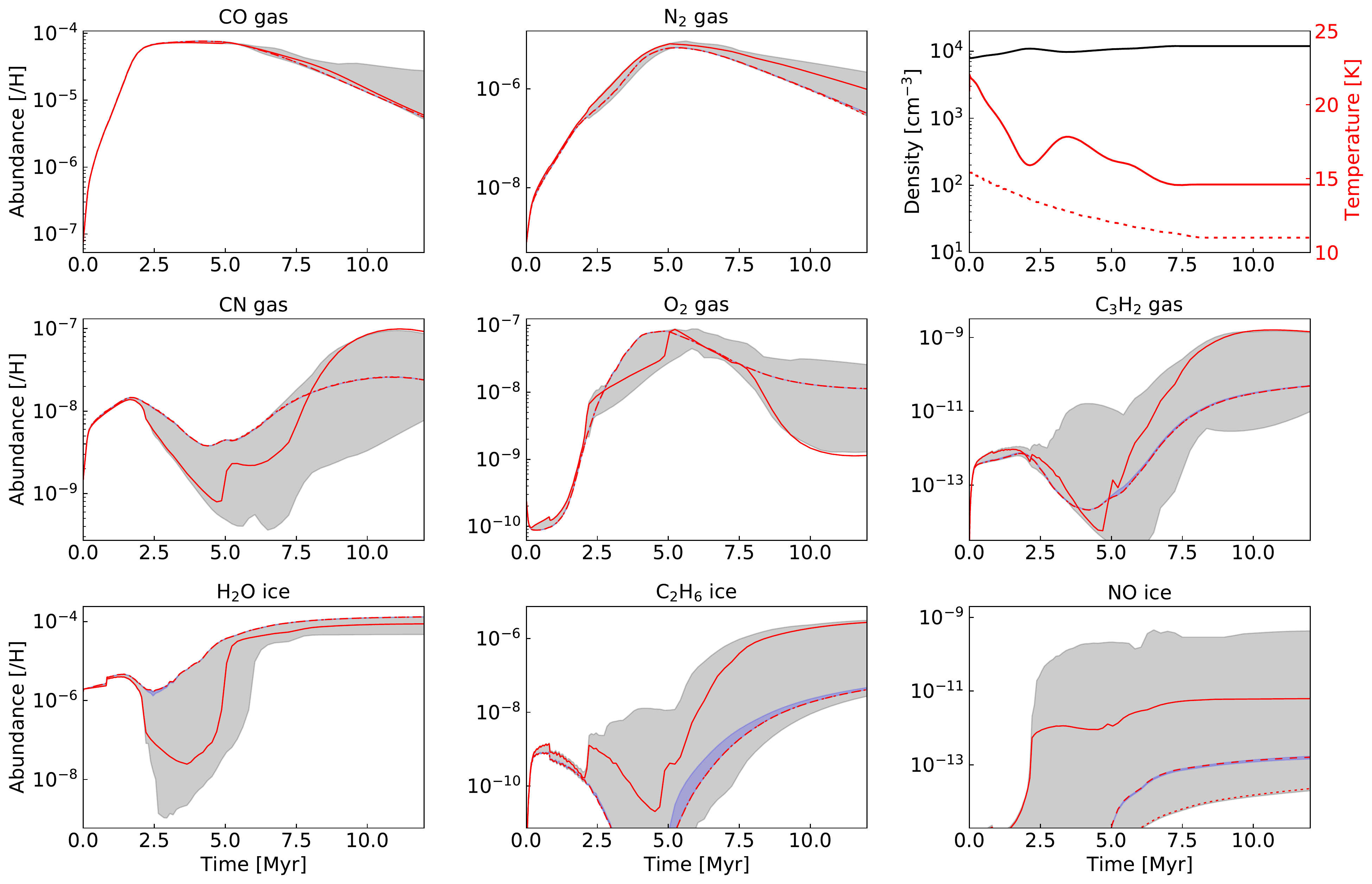}
    \caption{Top right panel) physical evolution in model MC. Other panels) abundance ranges for selected species with respect to H nuclei as functions of time in 10,000 different simulation runs, where the $f$ value for each species was randomly varied (gray shaded area).
    Purple shaded area indicates abundance ranges in another 10,000 simulations, where $f$ for several species was fixed to be 0.45 (see the main text).
    Red solid, dashed, and dotted lines show the results of models where $f$ was set to 0.2, 0.45, and 0.7, respectively, for all species except atomic H, \ce{H2}, and CO.
     }
    \label{fig:cl_models}
\end{figure*}

For the physical model of model MC, we adopt a one dimensional shock model
for the formation and evolution of a molecular cloud due to the compression of diffuse 
atomic gas by supersonic accretion flows \citep{bergin04}.
Throughout most of the simulation time, the density and temperature of the cloud are $\sim$10$^4$ cm$^{-3}$ and 10--15 K, respectively (top--right panel in Figure \ref{fig:cl_models}).
The column density of the cloud increases linearly with time, and the time required for the column density to reach $A_V$ = 1 mag is $\sim$4 Myr.
For model PE, we adopt a one-dimensional radiation hydrodynamics simulations for the evolution of a prestellar core to form a protostar via gravitational collapse \citep{masunaga00}.
Trajectories of fluid parcels are traced in the hydrodynamics simulation. 
Chemical evolution is solved along one chosen trajectory, 
in which the temperature increases from $\sim$10 K to $>$100 K, and the density increases from $\sim$10$^4$ cm$^{-3}$ to $\sim$10$^7$ cm$^{-3}$ (top right panel in Figure \ref{fig:hc_models}).
Additional details are provided in the Appendix B and in \citet{furuya15,furuya16}.

The chemical evolution of gas and ice are solved along the physical evolution.
We use a rate equation method, adopting a three-phase model \citep{hasegawa93} in which three distinct phases are considered: the gas-phase, a surface of ice, and the bulk of ice mantle.
Our chemical reaction network is based on that of \citet{garrod13}, with some minor updates \citep[see][]{aikawa20}.
The model takes into account gas-phase chemistry, 
interactions between gas and (icy) grain surfaces, and grain surface chemistry.
The top four monolayers of ice are assumed to be chemically active; 
the rest of the ice mantles is assumed to be chemically inert.
Two-body surface reactions occur solely via the Langmuir-–Hinshelwood mechanism and are treated using the modified rate equation method of \citet[][method A]{garrod08}.
As non-thermal desorption processes, we include photodesorption \citep[e.g.,][]{oberg07}, 
chemical desorption \citep[e.g.,][]{dulieu13,oba18}, 
and stochastic heating by cosmic rays \citep{hasegawa93b}.
In total, the network comprises 664 gaseous species, 281 icy species, and $\sim$9700 reactions, of which $\sim$1200 reactions are two-body reactions on surfaces.

A set of $\edes$ is taken from \citet{garrod13} with some updates \citep{wakelam17,shimonishi18}.
In particular, $\edes$ for atomic H, \ce{H2}, and CO are set to 440 K, 440 K, and 870 K, respectively.
To avoid an unphysically high abundance of \ce{H2} on grain surfaces, $\edes$ for \ce{H2} is decreased with increasing surface coverage of \ce{H2}, according to the method of \citet{garrod11} \citep[see also][]{penteado17}.
$\esd$ for CO is fixed at 240 K \citep{kouchi21}, 
while $\esd$ for atomic H is fixed to 220 K \citep{hama12}.
$\esd$ for \ce{H2} is assumed to be the same as that for atomic H.
For other species, $\esd$ is calculated by $f\edes$, 
where the value of $f$ is randomly assigned in the range 0.2--0.7 for each surface species.
Pre-exponential factors for desorption and diffusion rates are calculated by 
$\nu_i = \sqrt{2N_s \edes/(\pi^2 m_i)}$, where $N_s$ is the site density and $m_i$ is the mass of species $i$ \citep{hasegawa92}.
The equation can underestimate the pre-exponential factor (for desorption) by several orders of magnitudes, in particular for large molecules \citep[e.g., \ce{CH3OH}; see][and references therein]{minissale22}.
In addition, the pre-exponential factor for diffusion was not constrained by the UV-TEM experiments, 
and it may differ from that for desorption.
However, because we vary $\esd$, and the diffusion rate depends exponentially on $\esd$, the uncertainty in $\nu_i$ would be less important than that in $\esd$.


In all the simulations, the cosmic--ray ionization rate for \ce{H2} is set to $1.3\times10^{-17}$ s$^{-1}$.
Elemental abundances are taken from \citet{aikawa99} (see their Table 1).
In model MC, all the elements are initially assumed to be in the form of atoms or atomic ions.
In model PE, the initial abundances at the onset of collapse 
are adopted from those when $A_V$ reaches 2 mag (i.e., at 8 Myr) in model MC with a constant $f$ case with $f=0.45$ (see below);
the dominant forms of carbon, nitrogen, and oxygen are molecules (e.g., CO, \ce{H2O} ice, and \ce{N2}) rather than atoms \citep{furuya15}. 


\subsection{Results}
Gray shaded areas in Figure \ref{fig:cl_models} show the abundance ranges for selected species with respect to H nuclei as functions of time in 10,000 simulations of model MC.
In general, molecules that formed predominantly in the gas phase, such as CO and \ce{N2}, show small abundance ranges,
wheres molecules that formed on grain surfaces show larger abundance ranges \citep[see also][]{iqbal18}.
For comparison, we also ran simulations where $f = 0.2$, 0.45, or 0.7 for all species except for atomic H, \ce{H2}, and CO (called constant $f$ case), shown by red lines in Figure \ref{fig:cl_models}.
The constant $f$ case tends to underestimate the uncertainties in the abundances compared with the varied $f$ case, typically by a factor of $<$10.
Given the exponential dependence of the diffusion rates on $\esd$ (Equation \ref{eq:ksd}), the difference between the two cases might be smaller than expected.
The relatively small difference is attributable to the low dust temperature in model MC ($\sim$10 K), where hydrogenation reactions of atoms/molecules are efficient and other type of reactions, (e.g., radical-radical reactions) are less important.
We can define the critical value of $\esd$.
If $\esd$ is larger than the critical value, the actual value of $\esd$ is no longer relevant, because surface diffusion is too slow to affect molecular abundances in $\sim$10 Myr.
The critical value of $\esd$ is $\sim$500 K; 
thus species with $\edes$ greater than $\sim$2500 K 
(500 K divided by the minimum $f$ value of 0.2) are not relevant with the uncertainties.
In other words, species with $\edes$ lower than $\sim$2500 K
should dominate the uncertainties.
To identify key species that dominate the uncertainties in the abundance in the varied $f$ case, we calculate Pearson correlation coefficient between $f$ and the logarithm of the abundance at a given time as in \citet{penteado17,iqbal18}:
\begin{align}
P(f(i), \log(x(j, t))) =  \frac{{\rm cov}(f(i), \log(x(j,t)))}{\sigma(f(i))\sigma(\log(x(j,t)))},
\end{align}
where $x(j, t)$ is the abundance of species $j$ at given time $t$.
Based on the Pearson correlation, we identify the following key species with $|P| > 0.3$ for several species in the chemical network: atomic O ($\edes = 1320$ K), atomic N (720 K), HCO (2400 K), \ce{CH2} (1400 K), \ce{CH3} (1600 K), and \ce{NH} (1600 K).
When $f$ for the key species are fixed, the uncertainties in the abundance ranges are almost eliminated, which we confirmed by running another 10,000 simulations (blue shaded areas in Figure \ref{fig:cl_models}).

\begin{figure*}[t]
    \plotone{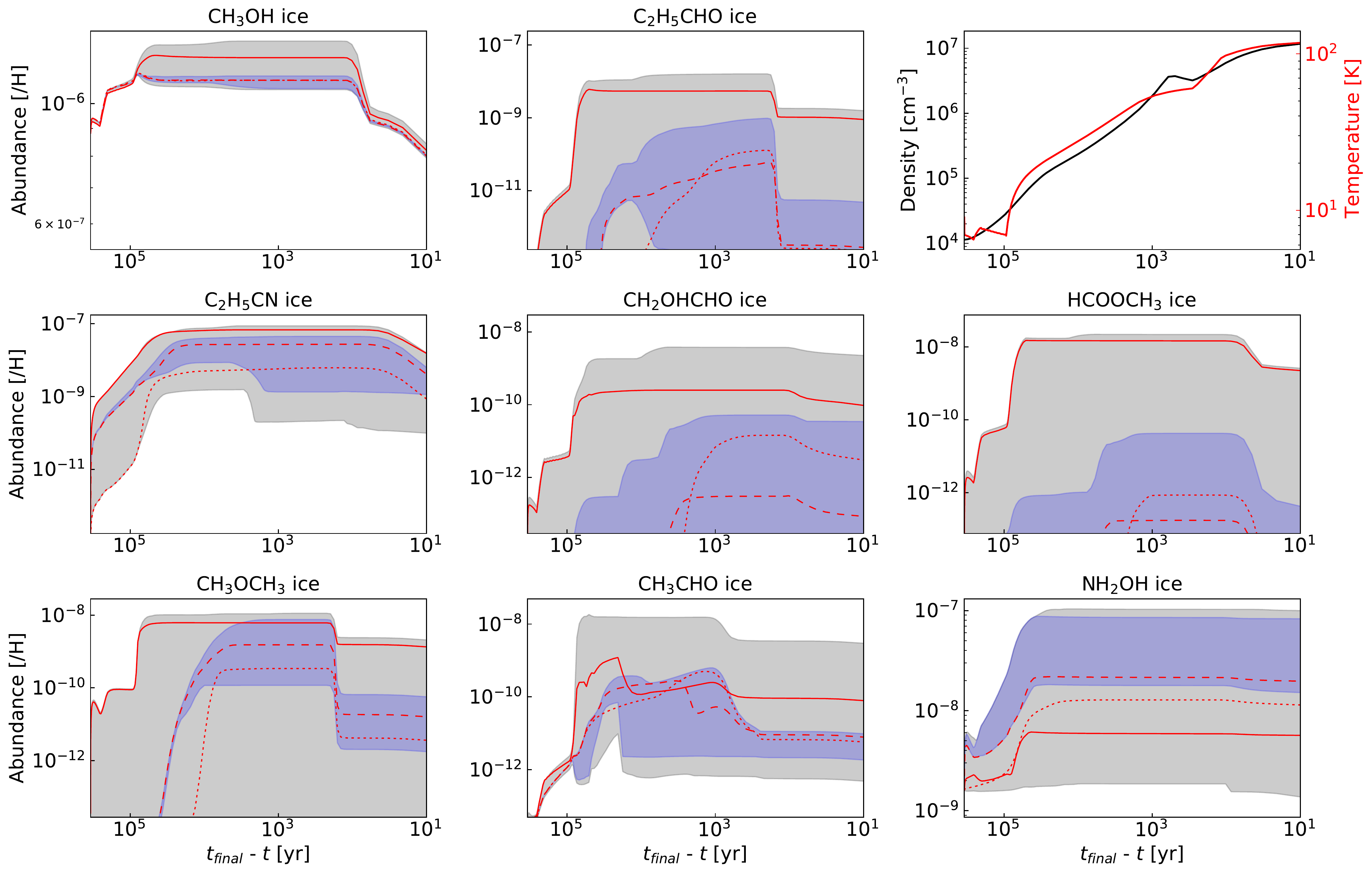}
    \caption{Similar to Figure \ref{fig:cl_models}, but for selected complex organic molecules in model PE. The horizontal axes represent $t_{\rm final} - t$, where $t_{\rm final}$ represents the final time of the simulation and $t = 0$ corresponds to the onset of the collapse to highlight the rapid changes of the physical conditions near and at the final stage.}
    \label{fig:hc_models}
\end{figure*}

The chemical composition of low-mass protostellar envelopes can be characterized by the abundances of complex organic molecules (COMs) and carbon-chain molecules \citep[e.g.,][]{sakai13,jorgensen20}. 
Figure \ref{fig:hc_models} is similar to Figure \ref{fig:cl_models} but shows the abundances of selected COMs in model PE (see also Figures \ref{fig:hc_gas} and \ref{fig:hc_2}).
The abundance ranges predicted by the varied $f$ case are often much larger than those predicted by the constant $f$ case.
The results appear to reflect a more complex surface chemistry in model PE than in model MC; more species are mobile on the surface because of higher dust temperatures ($\sim$10--100 K), whereas the efficient sublimation of atomic H and \ce{H2} slows simple hydrogenation reactions in model PE.
As in model MC, we calculated the Pearson correlation coefficients to identify key species that govern the uncertainties in COM abundances in the varied $f$ case.
We found that numerous species affect the uncertainties.
To demonstrate this, we ran an additional 10,000 simulations, where $f$ for radicals related to the main component of the ISM ice (\ce{H2O}, \ce{CO2}, \ce{CH4}, \ce{NH3}, and \ce{CH3OH}) and some other species (atomic N, atomic O, \ce{O2}, and NO) were fixed to 0.45.
As shown in blue shaded areas in Figure \ref{fig:hc_models}, uncertainties in the COM abundances were reduced compared with those in the original model but were still large (orders of magnitude).
The abundance ranges for carbon-chain species are smaller than those for the COMs (Figure \ref{fig:hc_wccc} in the Appendix), because carbon-chain species are mainly formed by gas-phase reactions triggered by the sublimation of \ce{CH4} \citep[e.g.,][]{sakai13,aikawa20}.


\section{Summary}
We measured $\esd$ for \ce{CH4}, \ce{H2S}, OCS, \ce{CH3OH}, and \ce{CH3CN} on c-ASW using UV-TEM without employing models. 
Compiling the $\esd$ values measured in this work and $\edes$ reported in the literature, we showed that 
the diffusion-to-desorption activation energy ratio ($f = \esd/\edes$) on c-ASW depends on the adsorbed species and ranges from $\sim$0.2 to $\sim$0.7.
No universal value exists for $f$, in contrast to the usual assumption made in astrochemical models.
We ran astrochemical models of molecular clouds and protostellar envelopes, randomly varying $f$ for each species.
In the case of molecular clouds, we identified several key species whose $\esd$ most strongly affects the uncertainties of the model predictions.
In the case of protostellar envelopes, we found that uncertainties in $\esd$ for many species  contribute to the uncertainties in the COM abundances.
It should be noted, however, that recent studies have suggested that non-diffusive mechanisms in prestellar stages are more important for the formation of some types of COMs than the Langmuir–Hinshelwood mechanism during the warm-up phase after the formation of a protostar \citep{jin20}.
If this is the case, our model of protostellar envelopes would overestimate the uncertainties in the COM abundances.

\acknowledgments
We thank the anonymous referees for helpful comments that helped us to improve the manuscript.
This work was supported in part by JSPS KAKENHI grant numbers 20H05847 and 21K13967.

\begin{appendix}
\renewcommand{\thefigure}{A\arabic{figure}}
\setcounter{figure}{0}
\renewcommand{\thetable}{A\arabic{table}}
\setcounter{table}{0}

\section{Additional figures and tables for Section 2}
\begin{figure}[ht]
\plotone{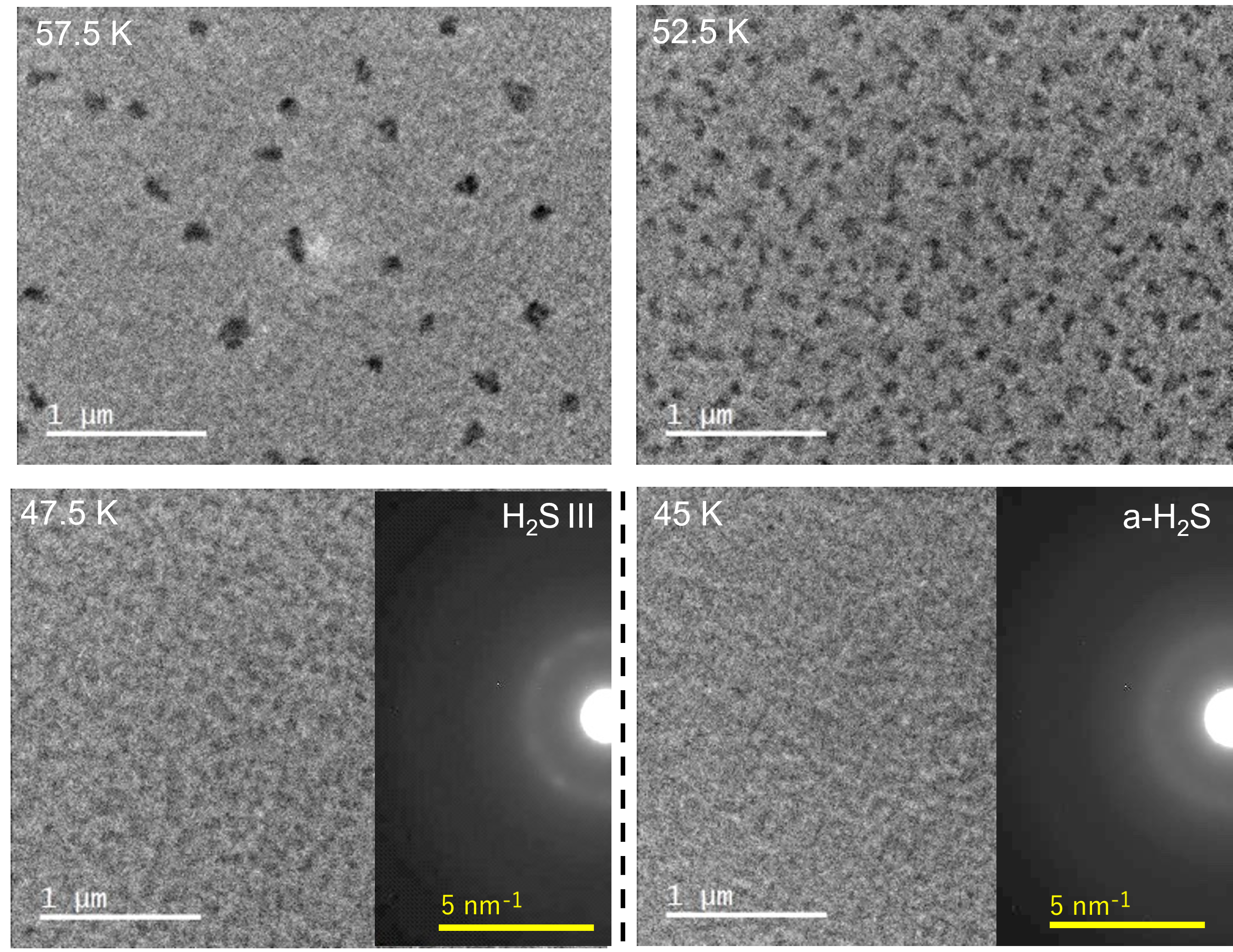}
\caption{
Similar to Figure \ref{fig:tem}, but for \ce{H2S}.
}
\label{fig:tem2}
\end{figure}

\begin{figure}[ht]
\plotone{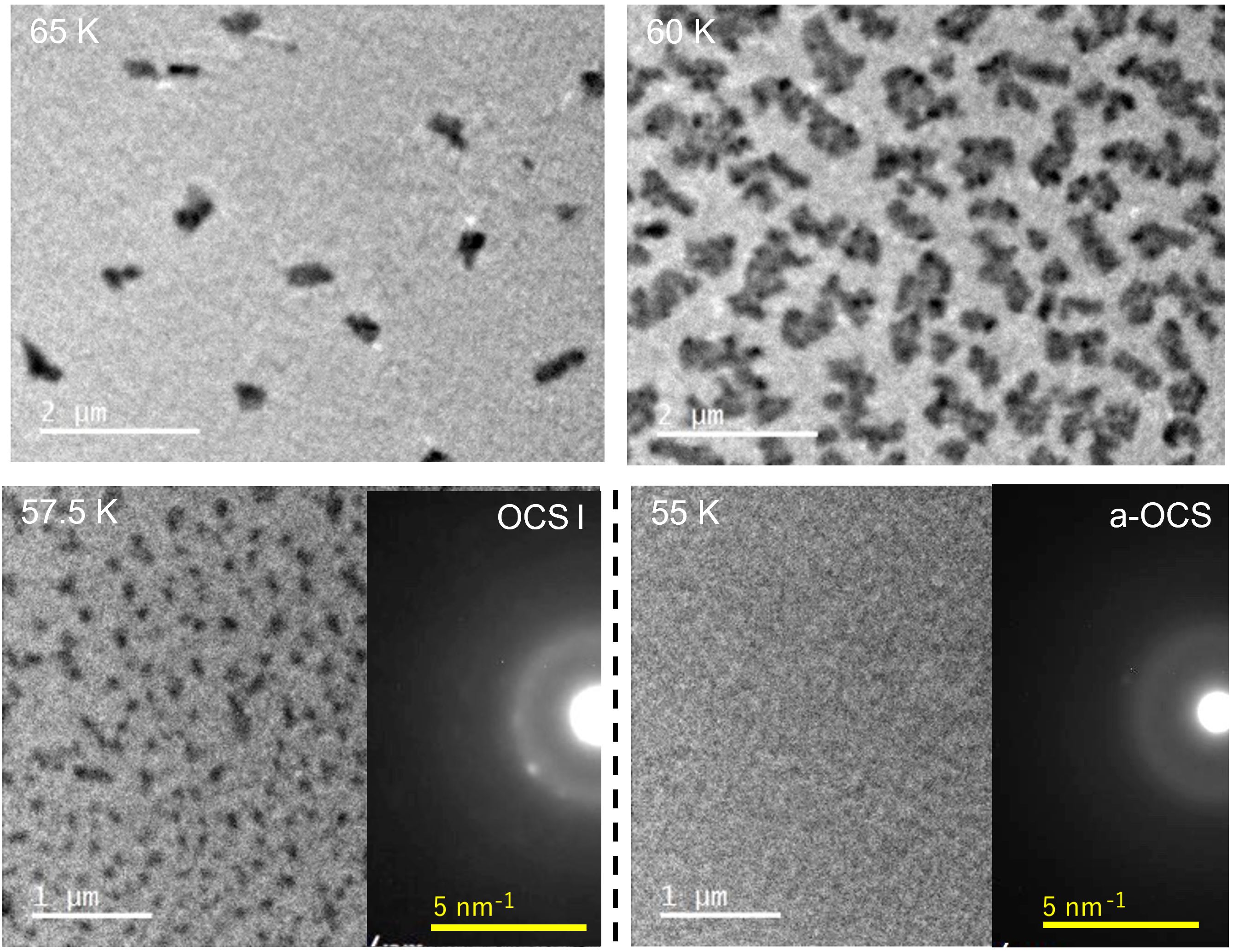}
\caption{
Similar to Figure \ref{fig:tem}, but for \ce{OCS}.
}
\label{fig:tem3}
\end{figure}

\begin{figure}[ht]
\plotone{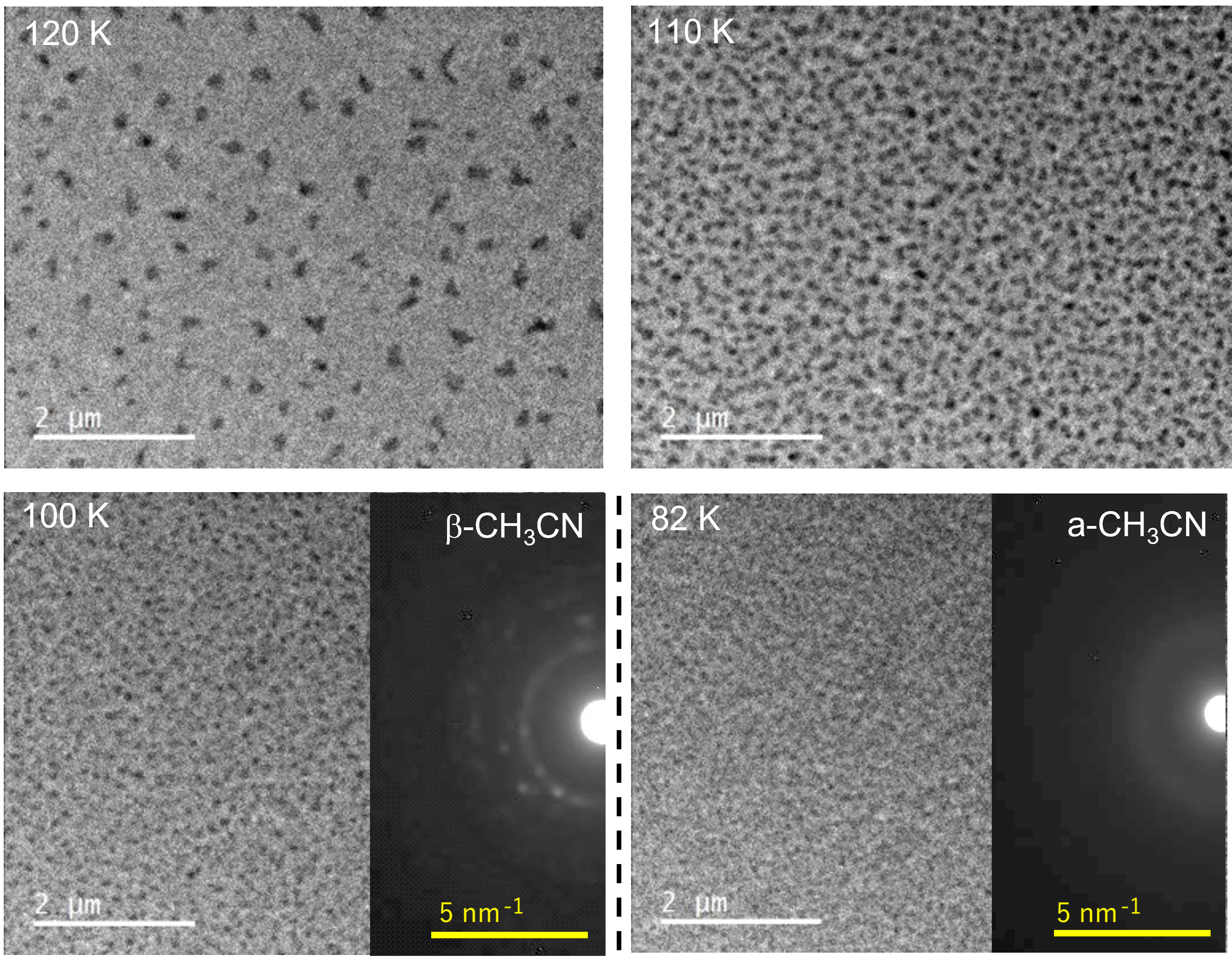}
\caption{
Similar to Figure \ref{fig:tem}, but for \ce{CH3CN}.
}
\label{fig:tem4}
\end{figure}

\begin{figure}[ht]
\plotone{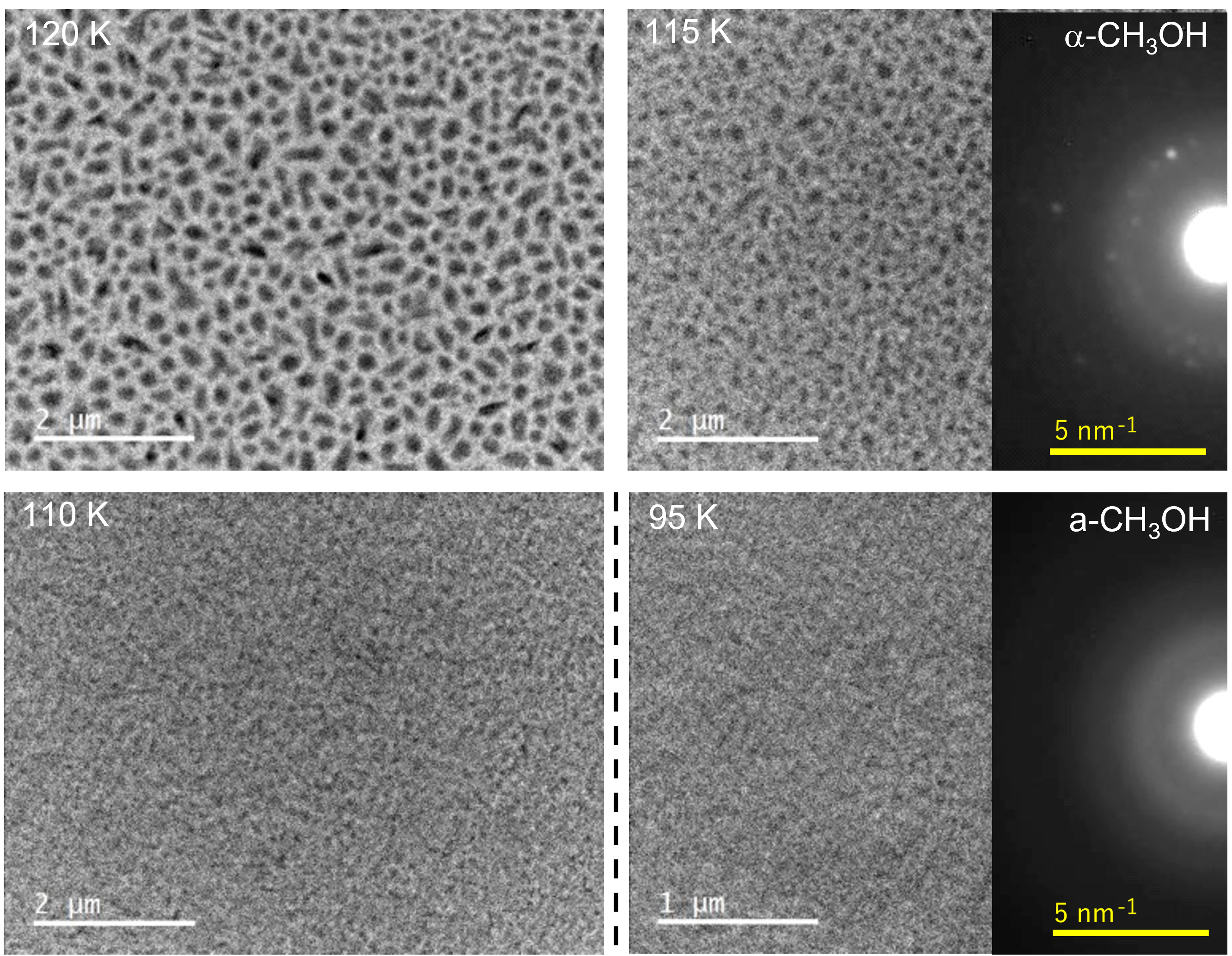}
\caption{
Similar to Figure \ref{fig:tem}, but for \ce{CH3OH}.
}
\label{fig:tem5}
\end{figure}

Figures \ref{fig:tem2}--\ref{fig:tem5} show TEM images after the deposition of \ce{H2S}, OCS, \ce{CH3CN}, and \ce{CH3OH} in the experiments presented in Section \ref{sec:experiment}.

$\esd$ measured with UV-TEM in this work and in our previous studies \citep{kouchi20,kouchi21} are summarized in Table \ref{table:esd_edes}, along with $\edes$ reported in the literature.
The values of $\esd$ were directly derived from the density of crystalline islands ($N_s$) without employing specific models as follows \citep[see also][]{kouchi20}.
Briefly, the mean distance between crystalline islands ($L$) were derived 
from the relation $L = (\pi N_s)^{-1/2}$, 
and the mean diffusion distance ($X$) is defined to be a half of $L$.
According to \citet[][Chapter 5.2]{smith95}, when desorption of molecules is ignored,
$X$ is expressed by
\begin{align}
X = a \left(\frac{N_s\nu_{sd}}{F}\right)^{\frac{1}{2}} \exp \left(-\frac{\esd}{2kT}\right), \label{eq:x}
\end{align}
where $a$ is hopping distance, $N_s$ is the number density of adsorption sites, 
$F$ is the deposition flux.
Then the slope on plot of $\ln X$ versus the inverse of the temperature corresponds to -$\esd/2kT$.

The impact of thermal desorption in our experiments is not significant.
The maximum substrate temperature in our experiments were 26 K for \ce{CH4}, 65 K for \ce{H2S}, 63 K for OCS, and 120 K for \ce{CH3CN} and \ce{CH3OH}.
These temperatures are slightly smaller or roughly correspond to the temperature where the molecules start to thermally desorb from ASW (i.e., only the molecules in the shallowest adsorption sites can desorb), but the molecules in most sites can remain on the ASW surfaces \citep[cf.][]{collings04,he16a}.
In addition, if the thermal desorption was significant, a slope of the plot of $\ln X$ versus the inverse of the temperature should be positive rather than negative.

In Equation \ref{eq:x}, only the interaction between \ce{H2O} and adsorbed molecules are considered, and the interaction between adsorbed molecules during the growth of islands is not considered.
In TEM observations, we could not obtain information on the adsorption of specific molecules on ASW.
In general, adsorption of molecules on the substrates are classified into partial adsorption, monolayer adsorption, and multilayer adsorption.
After adsorption, growth of islands or uniform film occurred depending on surface energy and interfacial energy.
However, in the most molecules used in this study, we have no information on the adsorption state and surface and interfacial energies.
Therefore, it is not straightforward to discuss the (possible) effect of the interaction between adsorbed molecules on the derived values of $\esd$.

\citet{he18} measured $\esd$ for \ce{CH4} and CO on ASW, employing infrared spectroscopy and Fick's law of diffusion.
They reported the $\esd$ value for CO and \ce{CH4} is $490 \pm 12$ K and $547 \pm 10$ K, respectively.
These values are higher than those measured with UV-TEM ($200 \pm 30$ K for \ce{CO} and $200 \pm 40$ K for \ce{CH4}).
The cause of the difference is unclear, because \citet{he18} and this work employed very different experimental methods.
The preparation method of ASW in \citet{he18} is different from that in this work;
ASW was annealed at 70 K for 30 min in \citet{he18}, while ASW was annealed at $\sim$100 K for 5 min in \citet{kouchi21} and in this work.
Because $\esd$ for CO on ASW deposited at 10 K without annealing is $350 \pm 50$ K as measured by \citet{kouchi20} using the UHV-TEM, the difference in the ASW preparation method may not fully explain the difference in the $\esd$ values derived in this work and in \citet{he18}.


\begin{table*}[ht]
\caption{Summary of our measured $\esd$ values and measured $\edes$ values reported in the literature}
\label{table:esd_edes}
\begin{center}
\begin{tabular}{cccccc}
\hline\hline
Molecule & Substrate & $\esd$ (K) & $\edes$ (K) & $\esd/\edes$ & Ref. \\
\hline
CO  & p-ASW  & $350 \pm 50$  & 980 & $0.36 \pm 0.05$& 1, 2\\
CO  & c-ASW  & $200 \pm 40$  & $849 \pm 55$ & $0.24 \pm 0.05$  & 3, 4\\
    &        &               & 870 & $0.23 \pm 0.05$  & 5\\
    &        &               & $1420 \pm 70$ & $0.14 \pm 0.03$  & 6\\
\ce{CH4}  & c-ASW  & $200 \pm 30$  & 1100 & $0.18 \pm 0.03$ & This work, 5\\
    &        &                     & $1370 \pm 70$ & $0.15 \pm 0.02$ &6\\
\ce{H2S}  & c-ASW  & $870 \pm 130$  & $2296 \pm 90$ & $0.38 \pm 0.06$ & This work, 7\\
\ce{OCS}  & c-ASW  & $1690 \pm 140$  & $2325 \pm 95$ & $0.73 \pm 0.06$ & This work, 7\\
\ce{CO2}  & c-ASW  & $1500 \pm 100$  & $2256 \pm 83$ & $0.66 \pm 0.04$ & 1, 4\\
    &        &                     & 2320 & $0.65 \pm 0.04$ & 5\\
\ce{CH3CN}  & c-ASW  & $670 \pm 210$  & $3790 \pm 130$ & $0.18 \pm 0.06$ & This work, 7\\
\ce{CH3OH}  & c-ASW  & $920 \pm 120$  & $3820 \pm 135$ & $0.24 \pm 0.03$ & This work, 7\\
\hline
\end{tabular}
\end{center}
\tablecomments{
(1) \citet{kouchi20}, 
(2) \citet{he16b},
(3) \citet{kouchi21},
(4) \citet{noble12},
(5) \citet{he16a},
(6) \citet{smith16},
(7) \citet{penteado17}
}
\end{table*}

\section{Physical models adopted for astrochemical models}
The physical model for model MC is based on the one-dimensional steady-state shock model of \citet{bergin04}.
This model mimics the scenario where molecular clouds are formed due to the compression of diffuse HI gas 
by super-sonic accretion flows \citep[e.g.,][]{inoue12}.
As time proceeds, the column density of post-shock materials (i.e., the molecular cloud) increases, 
which promotes molecular formation by attenuating the interstellar UV radiation.
The column density for post-shock materials at a given time $t$ after
passing through the shock front is $N_{\rm H} \approx 2\times10^{21}~(n_0/10~{\rm cm^{-3}})(v_0/15~{\rm km s^{-1}})(t/4~{\rm Myr})~{\rm cm^{-2}}$, 
where $n_0$ and $v_0$ are the preshock \mbox{\ion{H}{1}} gas density and velocity of the accretion flow, respectively.
$N_{\rm H}$ is converted into $A_V$ by $A_V/N_{\rm H} = 5\times 10^{-22}~{\rm mag/cm^{-2}}$.
The simulation is performed until $A_V$ reaches 3 mag (i.e., $\sim$12 Myr).
Throughout most of the simulation time, the density and temperature of the cloud are $\sim$10$^4$ cm$^{-3}$ and 10--15 K, respectively.

For model PE, we adopt the one-dimensional radiation hydrodynamics simulations of \citet{masunaga00}.
Initially, a prestellar core has an isothermal hydrostatic structure with a radius of $4\times 10^4$ au. 
The total mass of the core is greater than the critical mass for gravitational instability.
A protostar is born at the core center at $2.5\times10^5$ yr after the beginning of the gravitational collapse,
and the physical evolution is followed for an additional $9.3\times10^4$ yr 
(i.e., the total simulation time is $3.4\times10^5$ yr).
The trajectories of fluid parcels are traced in the hydrodynamics simulation.
We chose a fluid parcel, which is initially at a radius of 10$^4$ au and reaches 60 au from the protostar at the final time of the simulation,
and ran a gas-ice chemical astrochemical model along the stream line as in \citet{furuya16}.
Along the trajectory, the temperature increases from $\sim$10 K to $>$100 K,
while the density increases from $\sim$10$^4$ cm$^{-3}$ to $\sim$10$^7$ cm$^{-3}$ (top right panel in Figure \ref{fig:hc_models}).
According to \citet{furuya16,aikawa20}, who studied the distribution of molecules in this protostellar core model, the molecular abundances are mostly constant at $\lesssim$100 AU from the protostar.

\section{Additional figures for Section 3}
\begin{figure}[ht]
\plotone{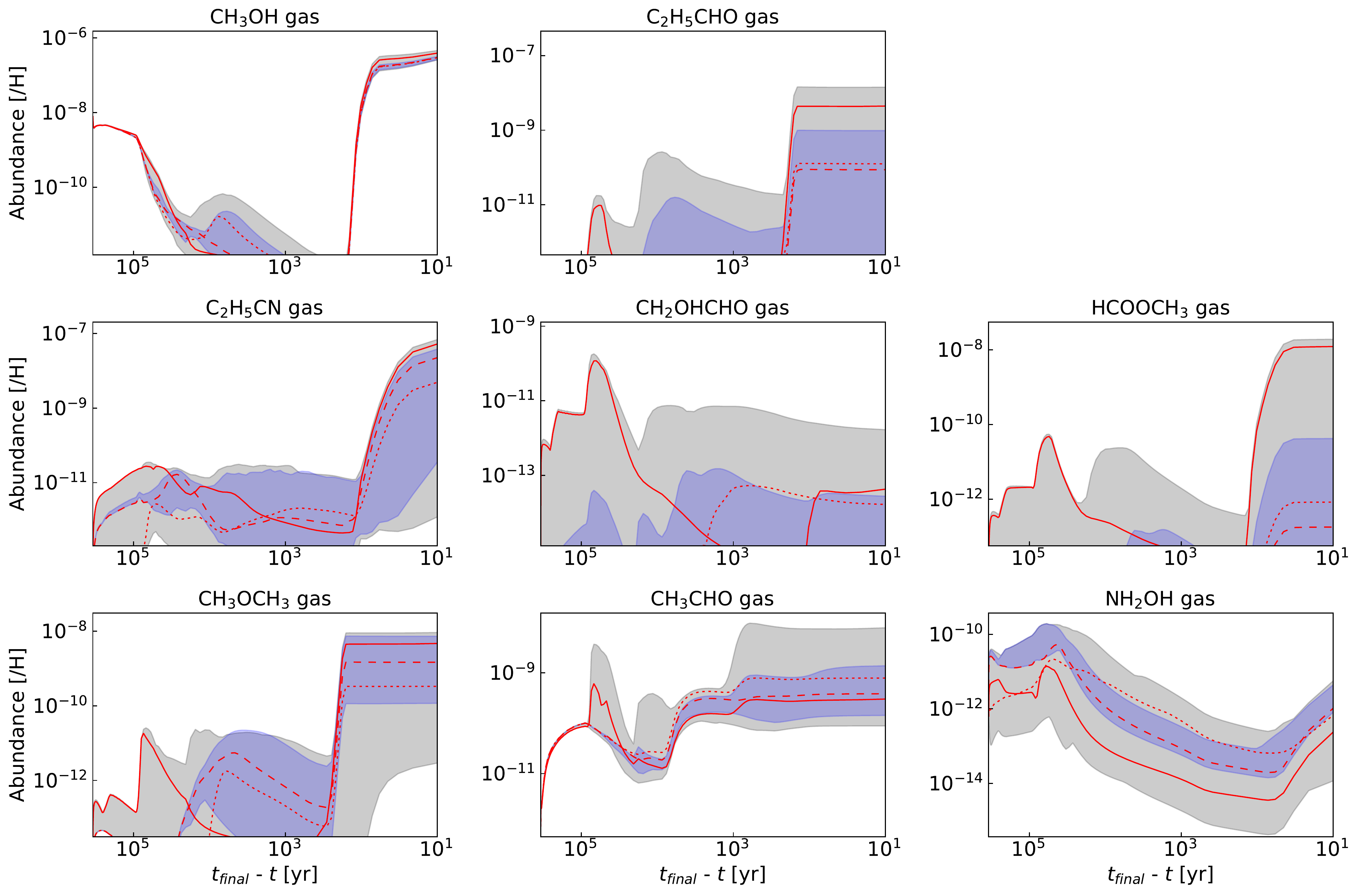}
\caption{
Similar to Figure \ref{fig:hc_models}, but for corresponding gas-phase molecules.
}
\label{fig:hc_gas}
\end{figure}

\begin{figure}[ht]
\plotone{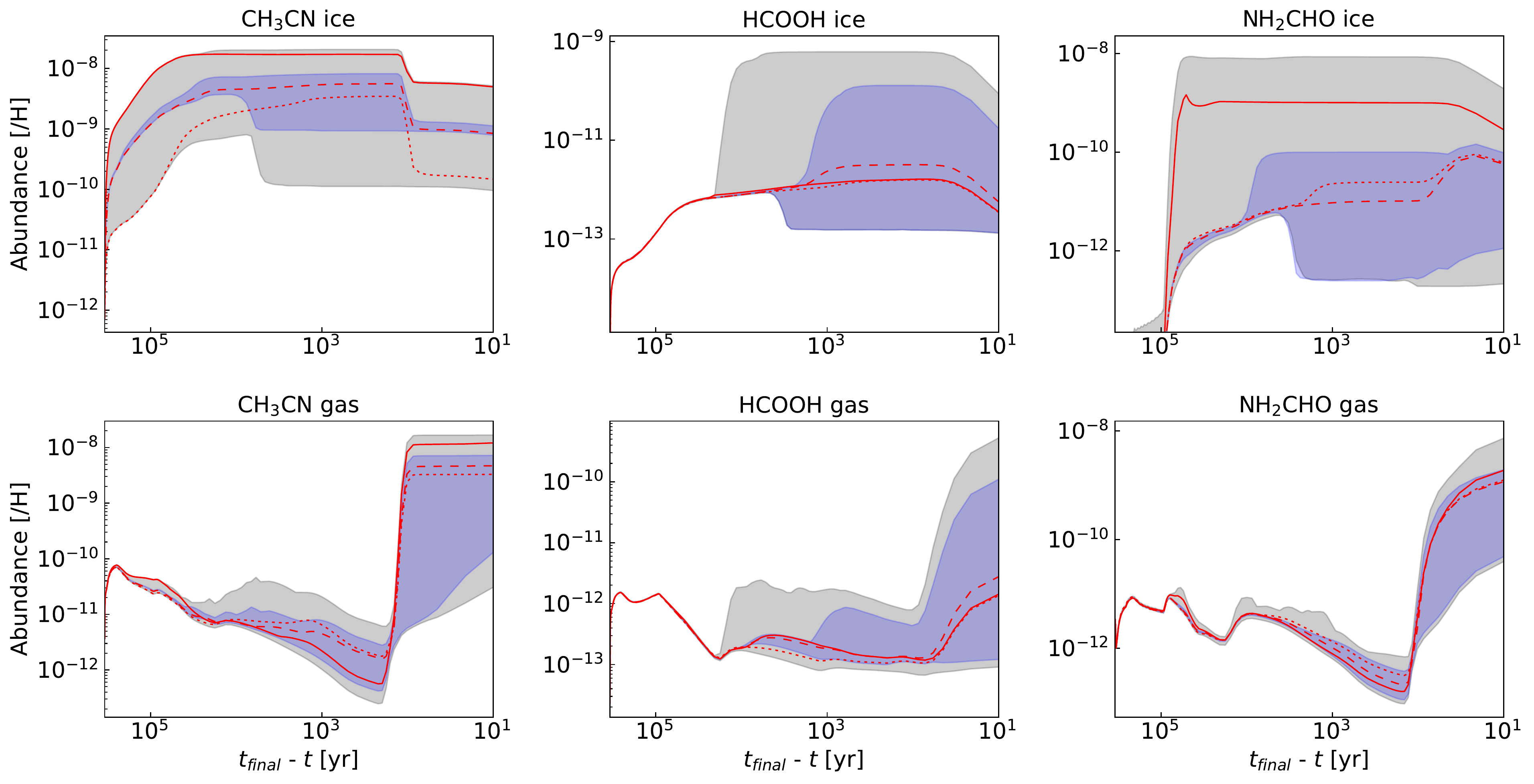}
\caption{
Similar to Figure \ref{fig:hc_models}, but for additional complex organic molecules in the gas phase and on dust grains.
}
\label{fig:hc_2}
\end{figure}

\begin{figure}[ht]
\plotone{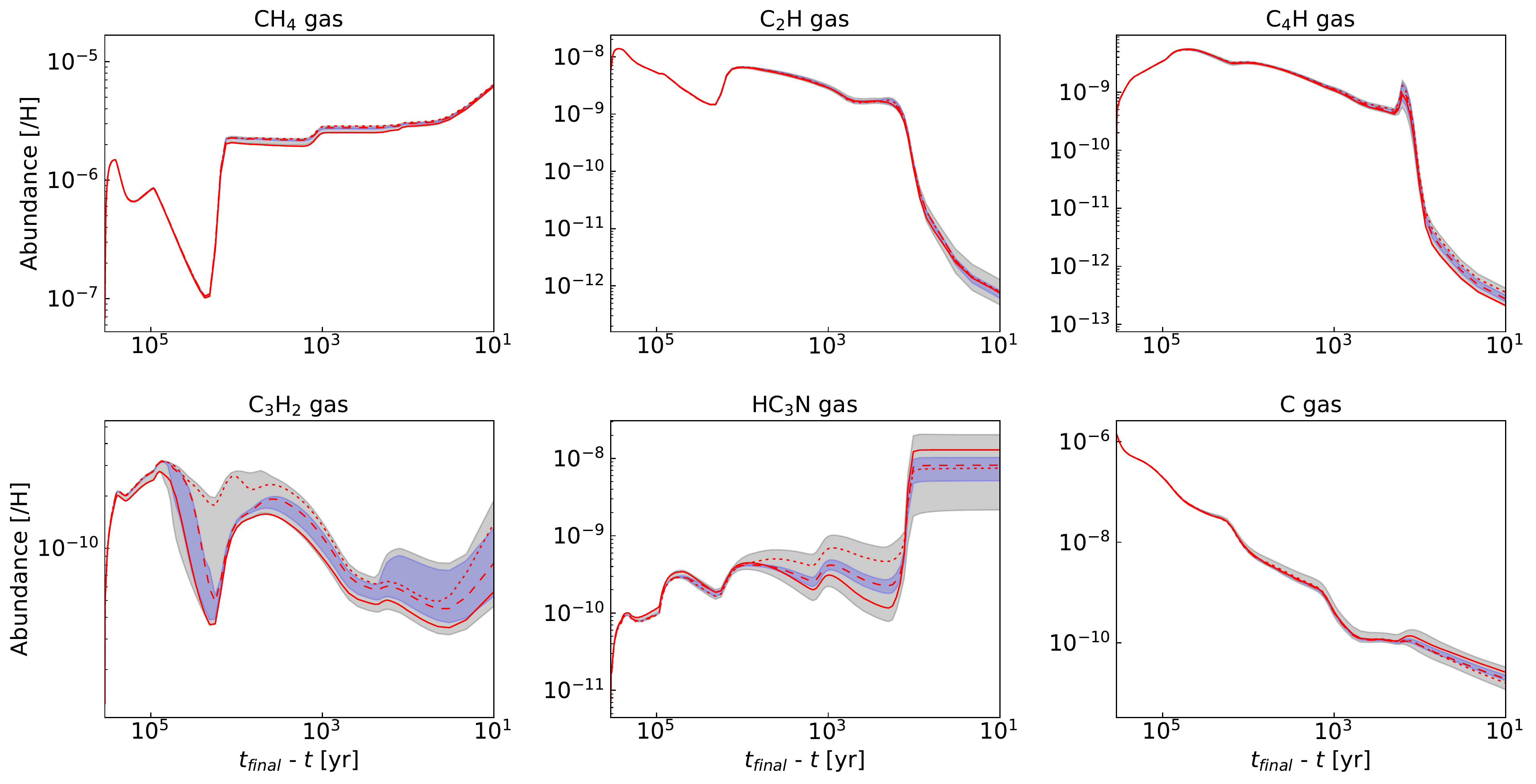}
\caption{
Similar to Figure \ref{fig:hc_models}, but for simple species and species relevant to warm-carbon-chain chemistry.
}
\label{fig:hc_wccc}
\end{figure}

Figures \ref{fig:hc_gas}--\ref{fig:hc_wccc} show the additional results of model PE for complex organic molecules and species relevant to warm-carbon-chain chemistry.

\end{appendix}

\bibliography{ms}{}
\bibliographystyle{aasjournal}






\end{document}